# Anomalous Electrical Conduction and Negative Temperature Coefficient of Resistance in Nanostructured Gold Resistive Switching Films

M. Mirigliano1, S. Radice1, A. Falqui2, A. Casu2, F. Cavaliere1, P. Milani1*

1CIMAINA and Department of Physics, Università degli Studi di Milano, via Celoria 16, 20133 Milano, Italy

2King Abdullah University of Science and Technology (KAUST), Biological and Environmental Sciences and Engineering (BESE) Division, NABLA Lab, 23955-6900 Thuwal, Saudi Arabia

Abstract:

We report on the observation of non-metallic electrical conduction, resistive switching, and a negative temperature coefficient of resistance in cluster-assembled nanostructured gold films above the electrical percolation and in strong-coupling regime, from room to cryogenic temperatures (24K). The structure of the films is characterized by an extremely high density of randomly oriented crystalline nanodomains, separated by grain boundaries. The observed behavior can be explained by considering space charge limited conduction and Coulomb blockade phenomena highlighting the influence of the high density of defects and grain boundaries on the localization of conduction electrons. Our findings have implications for a broad class of resistive switching systems based on random assemblies of nanoobjects.

*corresponding author: paolo.milani@mi.infn.it

The electric transport properties of granular metallic films (GMFs) differ drastically from their polycrystalline bulk counterparts: a significant departure from metallic conductivity and, in many cases, a negative temperature coefficient of resistance (TCR) are reported depending on their structure and composition [1–4]. GMFs consist of metal clusters or nanoparticles varying in size and structure, separated by a dielectric matrix (either vacuum or a non-conducting material) and characterized by a distinct electronic structure [5–7]. The electrical resistance of GMFs is strongly dependent on the coupling between adjacent units and has been studied by varying their density from very diluted (weak-coupling regime) to particle structural percolation (strong-coupling regime) [7–9]. Systems in weak-coupling regime have received particular attention in order to understand the role of defects and discontinuities in determining the non-metallic behavior, whereas systems in strong-coupling regime are reported to be ohmic with conventional transport mechanisms typical of polycrystalline metallic films [10–13].

Random networks of metallic nanowires/nanoparticles in a polymeric matrix, passivated by shell of ligands or oxide layers, have been used for the fabrication of non-linear circuital elements such as memristors and resistive switching devices for analog computing and neuromorphic data processing [14–21]. Recently we showed that continuous cluster-assembled gold films produced by the assembling of unprotected clusters, i.e. with no insulating matrix, also show resistive switching [18, 22, 23]. Their structure is characterized by the random stacking of differently shaped crystalline clusters directly connected by junctions of different cross sections with an extremely high number of defects and grain boundaries [22, 23].

Here we report that continuous cluster-assembled gold films, although in strong-coupling regime and above the electrical percolation threshold, show non-metallic electrical conduction and negative TCR within 24-300 K temperature range. The observed behavior indicates that conduction mechanisms typical of insulators or highly disordered semiconductors are occurring. Remarkably, the resistive switching activity of these systems is maintained down to cryogenic temperatures.

Nanostructured Au films are produced by a supersonic cluster beam deposition apparatus equipped with a Pulsed Microplasma Cluster Source (PMCS) [24], as described in detail in [22]. Au clusters with a bimodal log-normal mass distribution peaked around 5 nm are formed in an Argon atmosphere after the plasma ablation of a gold target [22, 25]; the cluster-Argon mixture is then expanded into a vacuum to form a supersonic beam directed on a silicon substrate with a thermally grown oxide layer [22, 23]. Clusters are deposited between two gold electrodes previously fabricated by thermal evaporation (Figure 1a). The amount of deposited material is measured by a pre-calibrated quartz microbalance, the evolution of the electrical resistance of the films is monitored *in situ* and *ex situ* in a two-probe configuration.

*In situ* electrical characterization in a range from room temperature (RT, 300 K) down to 24 K has been performed in vacuum ($10^{-5}$ mbar) on films mounted on a copper cold finger of a helium mechanical cryocooler. The film overall features have been investigated by high resolution TEM (HRTEM), using a spherical aberration-corrected microscope with an ultimate point resolution of 0.07 nm [26].

We characterized continuous films with an average thickness ranging from 15 nm to 30 nm, and resistance, before switching activation [19, 22, 23], varying from 80 Ω to 1000 Ω. This range belongs to strong-coupling regime where ohmic behavior and positive TCR is usually observed in granular films [7, 11, 27]. The value of the dimensionless tunnel conductance $g = h/4e^2 R_t(T \to \infty)$, where $R_t(T \to \infty)$ is the average tunnel resistance of the granular system at high temperature, discriminates the weak-coupling regime ($g < 1$) from the strong-coupling one ($g > 1$) [3, 7]. In Figure 1 b (left panel) we show the evolution of $g$, obtained by approximating $R_t$ to the resistance at RT, after the onset of percolation threshold, as function of the cluster-assembled film thickness [22]. The fast increase of $g$ is due to the high rate of deposition and the formation of connections among the clusters that open new conductive paths. Figure 1b shows the transition to the strong-coupling regime at a thickness value of roughly 7 nm.

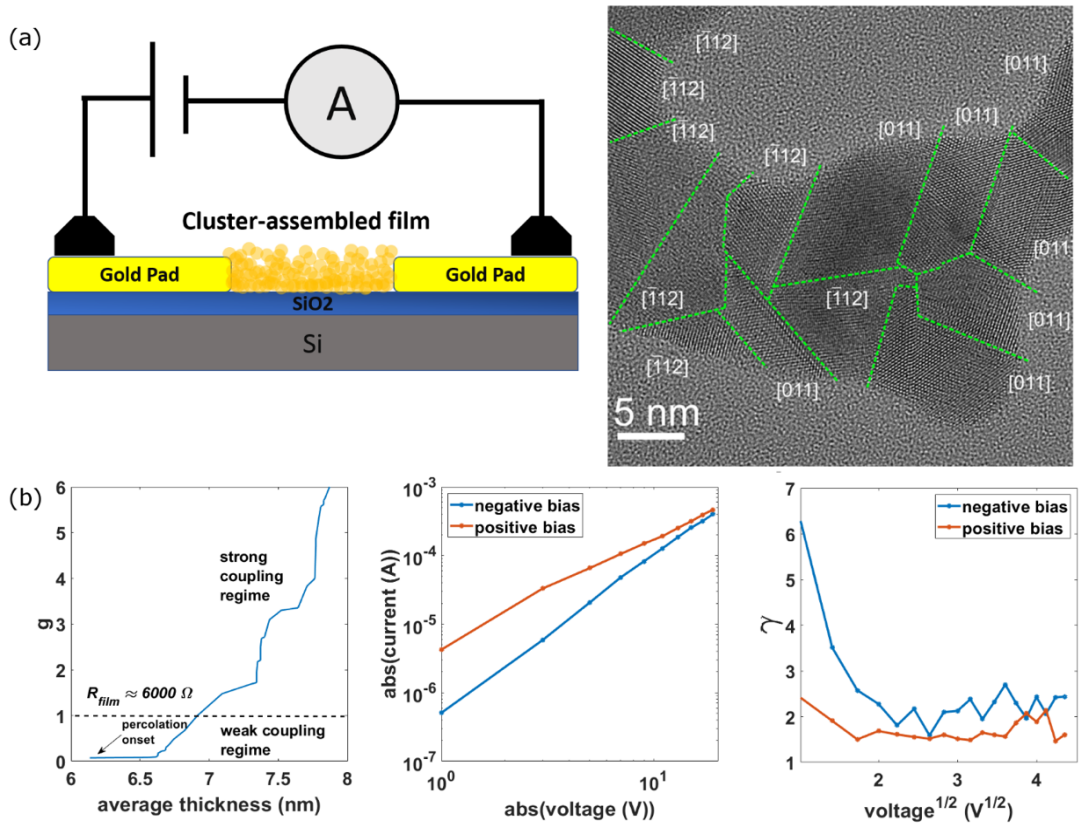

Figure 1a: Schematic view (not to scale) of the two-terminal device constituted by two thermally deposited gold electrodes bridged by a cluster-assembled film. The blue region beneath the gold films is a silicon oxide

layer. Electrical characterization is performed with an amperemeter connected in series to a voltage source at room temperature. On the top right, a HRTEM image of a typical region of the film, in submonolayer regime, is shown, where the different crystal domains constituting the film are separated by grain boundaries. For each single crystal domain, the corresponding zone axis [hkl] is displayed, after determination by local 2D-Fourier analysis of the relevant region of interest in the HRTEM image. Fig. 1b Left panel: evolution of the dimensionless tunnel conductance g after the percolation threshold, as function of the thickness of the film. Central panel: I-V curves in double logarithmic scale measured at RT. Red curve under positive bias, the blue one under negative bias. Right panel: $\gamma$ vs. the voltage square root.

Figure 1b (central panel) shows the room temperature I-V characteristics of the cluster-assembled film under positive and negative bias voltage in double logarithmic scale. A clear departure from an ohmic behavior is evident, with an asymmetry under positive and negative bias (red and blue curve respectively). The slope change observed in the curves is caused by the presence of switching events during the application of the voltage ramp [22, 23].

It is usually assumed that the electrical conduction properties of granular films in the strong-coupling regime are analogous to those of polycrystalline metallic films [3, 28]. The observed departure from an ohmic behavior at room temperature in continuous films resulting from the stacking of naked highly defective gold nanocrystals is unexpected. Our data suggest that different conduction mechanisms are taking place and dictated by the intimate structure of the films, which are characterized by an extremely high density of randomly oriented crystalline nanodomains, separated by grain boundaries and with a big amount of lattice defects (Figure 1a).

In the case of two-terminal devices based on semiconductor or insulating layers, information on the microscopic mechanisms determining the current-voltage characteristics can be extracted by considering the parameter $\gamma = \frac{d\ln(I)}{d\ln(V)}$, where $\ln(I)$ and $\ln(V)$ are the logarithms of the current and of the applied voltage, respectively [29, 30]. The analysis is carried out by plotting $\gamma$ against $V^{\frac{1}{2}}$, since this curve has a well-defined trend for different mechanisms such as ohmic, space charge limited conduction (SCLC), Schottky, Poole-Frenkel, tunnelling, etc. [7, 29].

The trend of the gamma parameter for our films is reported in Figure 1b (right panel) showing a transient (blue curve) before stabilizing around a value nearly 2, which is typical of SCLC [29, 31, 32]. In this regime, the free carrier density is low and the electrical conduction is usually determined by the charges injected from ohmic electrodes [29].

In systems characterized by SCLC, ohmic conduction is usually observed at low bias voltages, due to the presence of a small fraction of thermally generated carriers [31]. Cluster-assembled gold films show a non-linear I-V curve even at very low voltages (data not shown) suggesting that different concurrent mechanisms contribute to determine a SCLC regime and lower the free electron density. Coulomb blockade [9, 33] and defect localization effects [7, 13, 34] could be possible causes for the low concentration of free carriers in our systems, due to their overall disordered crystal structure, which manifests as a very high density of grain boundaries as observed by HRTEM (Figure 1a).

To gain a deeper insight about the phenomena involved in the conduction process of cluster-assembled Au films and to discern among different mechanisms, we investigated the evolution of electrical conduction with temperature [9, 35, 36].

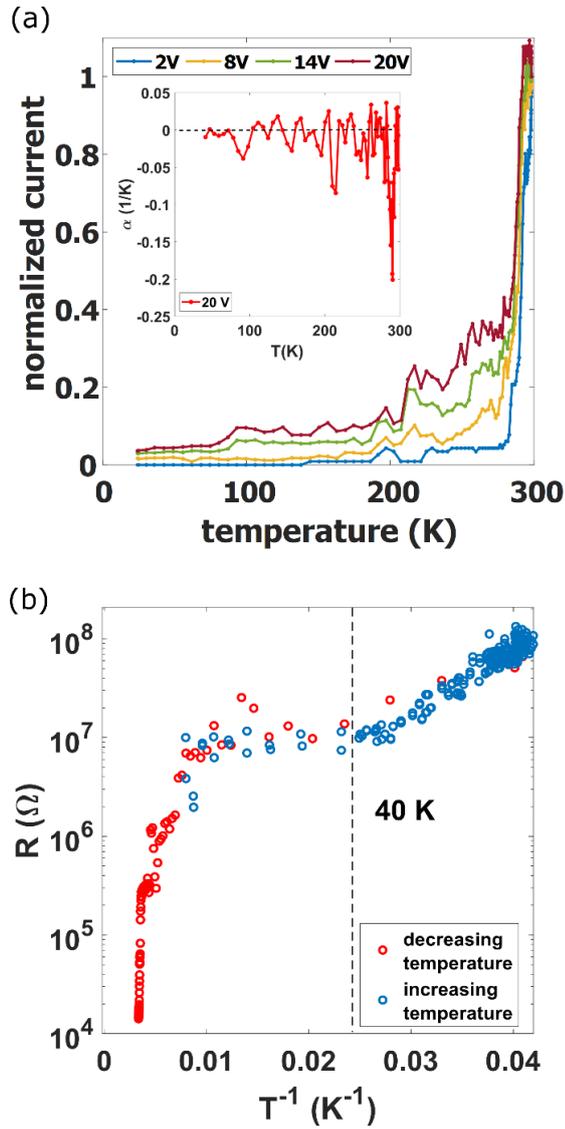

Figure 2a: Current normalized to the value measured at RT as function of the temperature for different applied voltages. The inset shows the temperature coefficient of resistivity (TCR) for the curve measured upon the application of 20 V. b: Resistance as a function of the inverse of the temperature, both for cooling (square point) and heating (cross) cycles, in logarithmic scale. An Arrhenius-like trend is recognizable only for temperatures below 40 K.

Figure 2a shows the temperature dependence of the current at different applied voltages normalized to the value measured at RT. We observe a steep decrease of the current in the range between RT and 250 K; from 250 K to 24 K the decrease continues with a lower slope. The trend of the current for different applied voltages is qualitatively similar.

In metallic systems, finite electrical resistivity arises due to scattering processes from impurities or various thermal excitations [6, 37–39]. The scattering events can be considered as statistically independent and thus additive, leading to the Matthiessen's rule, where any thermally induced scattering simply increases the resistivity $\varrho(T)$ [40, 41]. This corresponds to a positive Temperature Coefficient of Resistivity (TCR), i.e. $d\varrho/dT > 0$.

Cluster-assembled gold films are characterized by a trend of resistance with temperature typical of non-metallic systems. The inset of Figure 2a shows that cluster-assembled gold films have a negative TCR in particular near RT, the oscillation around zero at lower temperatures can be ascribed to the presence of switching events of amplitude smaller than the resistance temperature variations [22, 23].

These results are unexpected for a nanostructured metallic film in the strong-coupling regime: to the best of our knowledge only discontinuous gold ultrathin films and layers of molecularly linked gold nanoparticles have been reported to show non-metallic electrical conduction with temperature [2, 3, 12, 27], although not in such a large temperature range. Self-assembled films of CnS2-linked Au nanoparticles have electrical properties ranging from insulating to metallic-like depending on the separation of the Au building blocks [10]. In the insulating regime, electric transport occurs through cooperative electron tunneling (co-tunneling) at low temperatures, variable-range hopping (VRH) at intermediate temperatures, and Arrhenius-type behavior at high temperatures [34]. The weight of each of these contributions depends on both the interparticle separation and the spatial organization.

Similarly, discontinuous films composed by irregularly shaped gold islands, assembled by atom deposition, with density close to the percolation threshold [3, 11] show non-metallic transport strongly influenced by local disorder causing variations in the tunnel junction gaps and in the Coulomb blockade energies, due to island size fluctuations and offset charges [42]. A conduction percolation (co-percolation) model is applied to determine the total electrical current through the film as a function of both temperature and bias voltage [9]. The flowing of current is described as a percolation process through the ramified metallic islands. Unlike the case of hopping regime, in this case the high degree of disorder is related to the wide distribution, without mutual correlation, of the island electrostatic charging energy and of the parameters that characterize different tunnel junctions [3, 9]. Increasing the island density till the reaching of strong-coupling regime, an ohmic electric transport is observed [3, 9, 27].

Various types of disorder are considered at the origin of a negative TCR in ultrathin discontinuous films: i) fluctuations in the tunnel gaps between adjacent islands; ii) variations in the size and shape of the islands; iii) random offset induced by trapped impurity charges in the substrate [9, 40]. In our case, cluster-assembled films are continuous and in the strong-coupling regime, however they do not show the electrical behavior typical of continuous metallic films. We suggest that this is due the extremely high concentration of defects and grain boundaries slicing the crystal domains that constitute the films: upon deposition on the substrates, gold clusters formed in the gas phase do not lose their individuality and give rise to the multidomain structure, as confirmed by HRTEM analysis [23] (Fig. 1a). The overall morphology of the films can be described as constituted by grains and bridges interconnecting plenty of different aggregates. This kind of spatially extended disorder is substantially different from what observed in polycrystalline metallic films grown by atom deposition, where the density of grain boundaries is much lower compared to that we find in our systems [43].

Figure 2b displays the resistance vs the inverse of the temperature in logarithmic scale, showing that an Arrhenius-like behavior, characterized by a relation $R(T) \propto \exp\left(\frac{T_0}{T}\right)$, is not detected except for temperatures below 40 K. The observed behavior deviates from a pure hopping conduction like that in the Efros-Shklovskii model [7, 9, 44] for discontinuous films. On the other hand, the co-percolation model successfully describes the non-Arrhenius behavior of the electrical resistance at low bias voltage [3, 9], even if further type of disorder, such as Anderson localization, could contribute to the overall electrical behavior observed [34, 45]. We also note that the trend is reversible, i.e. the resistance curve obtained during the cooling coincides well with that obtained during the heating, strongly suggesting that the observed behavior is not related to a phase transition but only to electronic properties.

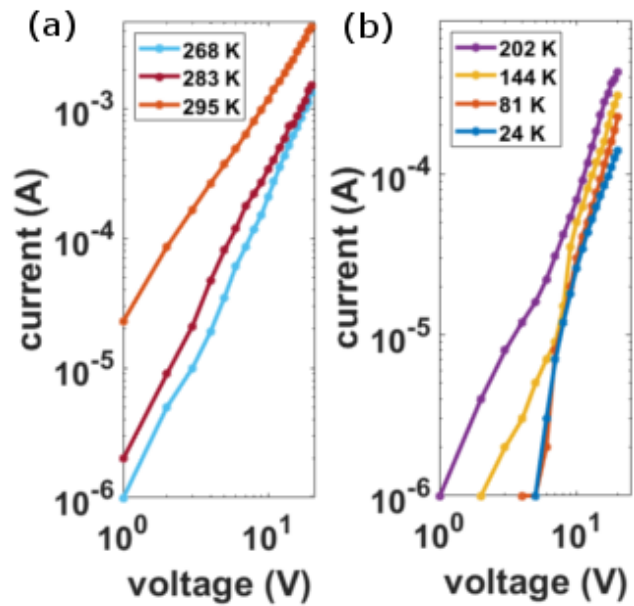

Figure 3: (a) I-V curves for different temperatures in the range 295 K to 268 K. (b) I-V curves in the range 202 K to 24 K.

The co-percolation model predicts not only the non-Arrhenius behavior but also a power law I-V characteristic (see Figure 1b). Figure 3a shows the I-V characteristics in the temperature range 295 K to 268 K. The trend is constantly linear in double logarithmic scale and unaffected by thermal change except for a higher resistance measured at lower temperatures. On the other hand, in Figure 3b the curves show steep slope variation for temperatures lower than 144 K. Although the trend slightly deviates from a pure power law, this agrees with the co-percolation model considering the occurrence of Coulomb blockade at cryogenic temperatures [9, 33].

At low temperatures we also notice that the $\gamma$ parameter explores values larger than 2. This agrees with the observation of higher resistance states at low temperatures and indicates that the variation of thermally generated carriers is at the origin of the effects observed at low voltages. In addition, we underline that the resulting parameter $\gamma$ is not compatible with a pure SCLC, but it reflects the dependence of the current by an external voltage in a medium with i) low density of free carriers and ii) electrostatic phenomena hampering charge flow.

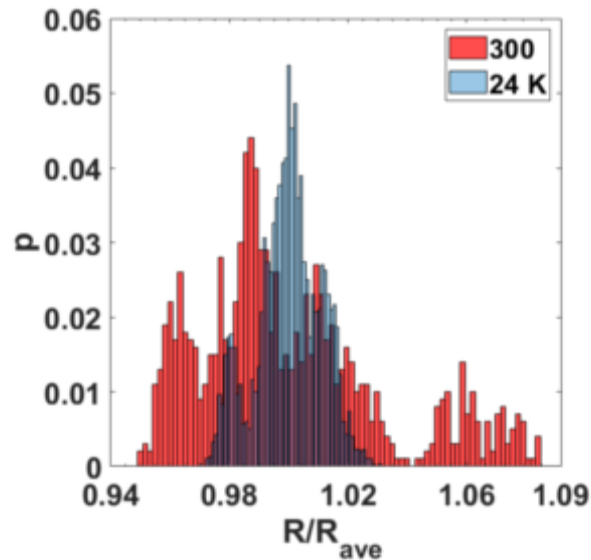

Figure 4: Distribution of the resistance values, of a 15 nm thick sample, normalized by their average for the measurements carried out at constant voltage (5 V) at RT (red data) and at 24 K (blue data).

Cluster-assembled gold films exhibit resistive switching (RS) phenomena, under the application of pulsed or continuous voltages [18, 19, 22, 23]. The evolution of the RS activity with temperature can provide useful elements for the understanding of the microscopic conduction mechanisms. In Figure 4 we show a typical histogram of the resistance, measured at both RT and 24

K upon the application of a constant voltage, for a duration of 1000 s. The different peaks in the histogram are due to the different resistance levels explored during the resistive switching phenomena. Remarkably we observe a substantial RS activity at 24K spanning a lower number of levels compared to that at RT. We can interpret these results by considering that the flow of electric current causes the rearrangement of domains grown randomly with their related lattice defectivity and grain boundaries. As a consequence, concomitant dynamical creation and destruction of pathways with variable resistance occur through the rearrangement of defects and grain boundaries [16, 18]. This process is favored at RT by the high mobility of atoms and atomic planes [46], while the latter is substantially reduced at cryogenic temperatures. Moreover, the persistence of RS events at cryogenic temperatures could be related to the structure of cluster-assembled films characterized by a landscape crowded of defects and interconnects between grains resulting in an assembly of interacting nanojunctions [42, 47, 48]. Multiple conductance states are observed in single metallic nanojunctions at cryogenic temperatures [49, 50] with electrical conductions characterized by discrete steps of conductance involving Coulomb blockade phenomena both in increasing and decreasing resistance [49, 51]. Cluster-assembled gold films can be then considered as an assembly of nanojunctions connected in series and in parallel, thus displaying a collective electrical behavior resulting in resistive switching phenomena [18, 23, 52].

In summary, nanostructured continuous Au films obtained by the assembling of nanocrystalline units, produced in the gas phase, exhibit a departure from ohmic behavior, resistive switching, and a negative TCR over a thermal range from RT to cryogenic temperatures. Our data can be explained by considering SCLC and Coulomb blockade phenomena, similarly to what observed in highly disordered semiconductor or insulator films. Of primary importance for the understanding of the microscopic mechanisms responsible for these puzzling electrical properties is the influence of an extremely high density of grain boundaries and lattice defectivity on conduction electron localization. These results highlight that cluster-assembled gold films are a challenging platform for exploring the fundamental role of extended nanoscale defects on electron localization and transport mechanisms, and for the fabrication of resistive switching devices that can operate over a wide temperature range for neuromorphic data processing [23, 53, 54].